\documentclass{sig-alternate-05-2015}

\begin{document}

\title{A Gaze-Assisted Multimodal Approach to Rich and Accessible Human-Computer Interaction}

\numberofauthors{1} 
\author{
  \alignauthor Vijay Rajanna, Dr. Tracy Hammond\\
    \affaddr{Sketch Recognition Lab. Dept. of Computer Science and Engineering}\\
    \affaddr{Texas A\&M University, College Station}\\
    \email{vijay.drajanna@gmail.com, thammond@gmail.com}\\
}


\maketitle
\begin{abstract}
Recent advancements in eye tracking technology are driving the adoption of gaze-assisted interaction as a rich and accessible human-computer interaction paradigm.
Gaze-assisted interaction serves as a contextual, non-invasive, and explicit control method for users without disabilities; for users with motor or speech impairments, text entry by gaze serves as the primary means of communication.
Despite significant advantages, gaze-assisted interaction is still not widely accepted because of its inherent limitations: 1) Midas touch, 2) low accuracy for mouse-like interactions, 3) need for repeated calibration, 4) visual fatigue with prolonged usage, 5) lower gaze typing speed, and so on.
This dissertation research proposes a gaze-assisted, multimodal, interaction paradigm, and related frameworks and their applications that effectively enable gaze-assisted interactions while addressing many of the current limitations.
In this regard, we present four systems that leverage gaze-assisted interaction: 1) a gaze- and foot-operated system for precise point-and-click interactions, 2) a dwell-free, foot-operated gaze typing system. 3) a gaze gesture-based authentication system, and 4) a gaze gesture-based interaction toolkit.
In addition, we also present the goals to be achieved, technical approach, and overall contributions of this dissertation research.
\end{abstract}

\section{Introduction}
\label{sec:intro}
Gaze tracking (eye tracking) refers to tracking and measuring eye movements to determine the point of gaze (target location)~\cite{GazeInteraction:Book:Majaranta}.
Real-time information of the eye movements gathered through eye trackers can be used for direct manipulation of interface elements; this forms the basis of gaze-assisted interaction \cite{GazeInteraction:Book:Majaranta}.
Hence, gaze-assisted interaction is quite promising as we explore interaction paradigms beyond the conventional mouse- and keyboard-based interaction methods.
Interfaces controlled with gaze not only support immersive interactions, but also serve as an accessible technology for users with physical impairments, who find it difficult to use conventional interaction methods \cite{visionenhancement}.
Besides enabling hands-free interactions, gaze-assisted interaction has an inherent advantage because of the way we interact with user interface elements on a computer.
For example, to click a button, we first look at it, move the cursor from its current location onto the button, and finally click it.
However, what if we can activate (click) the target when the user first looks at it.
This avoids a) switching the hand between the keyboard and mouse, and b) a lot of mouse movements on a big screen or multi monitors.
Hence, gaze interaction is specifically suitable for applications that make limited use of keyboard input but rely heavily on mouse input.
Some example use cases include web browsing, map interaction, reading tasks, and so on.

However, despite these advantages, gaze-assisted interaction suffers from the ``Midas Touch" issue \cite{Jacob:WYLWYG}, which can be summarized as the lack of efficient methods to execute a user's commands at the point of regard.
A command (e.g., click) can not be executed wherever a user looks on the screen, this is because, a user first scans the display before fixating on the point of regard to execute an action.
Hence, any gaze controlled interface that can not accurately distinguish between an intentional and unintentional visual focus is hardly suitable for practical applications.
Existing solutions use dwell time or a blink to execute a user's commands at the point of regard, but, both dwell- and blink-based activations are limited by accuracy, speed, and inability for a prolonged usage.
We believe that an effective solution, enabling gaze to be used to perform mouse-like interactions, should use gaze input only for pointing, but a supplemental input should be used to execute commands at the point of regard.
Hence, in this dissertation proposal, we present a multimodal interaction paradigm that effectively combines gaze and foot input modalities to achieve point-and-click interactions on a computer.
We discuss two systems, GAWSCHI and gaze typing, that utilizes this gaze and foot interaction paradigm.
Furthermore, we re-contextualize gaze input as gestural input to create gaze gesture-based interactions.
In this method, we use sketch recognition principles and pattern matching algorithms to translate gaze input into gestural input.
We discuss two systems, gaze authentication and a gaze gesture toolkit that utilize gaze gesture-based interaction paradigm.


\section{Prior Work}
The feasibility of gaze-assisted interaction was first demonstrated by Jacob~\cite{Jacob:WYLWYG}.
Since Jacob's work, gaze-assisted interaction has been further explored by combining gaze input with various input modalities to create muti-modal interaction methods.
Similar to gaze input, the design of foot-operated input devices has been a research focus for long time in the field of human-computer interaction.
A major work in foot input modality was presented by Pearson and Weiser~\cite{Pearson:Weiser}, who proposed the design of ``moles," which are foot-operated input devices similar to a mouse.
Pakkanen and Raisamo~\cite{Pakkanen:spatial}, demonstrated how foot input can be leveraged for executing non-accurate spatial tasks.
However, only G\"{o}bel et al. \cite{gaze:foot:Gobel}, have demonstrated a multi-modal interaction method by combining gaze and foot input for secondary tasks like pan and zoom interactions.
Focusing on some of the major research in gaze gesture-based interaction, Drewes et al.~\cite{Drewes2007}, presented a framework to interact with computers using eye gaze.
The authors implemented a gaze gesture algorithm based on mouse gestures, where users move their gaze in a combination out of eight directions to draw a gesture and execute an action. 
Wobbrock et al.~\cite{Wobbrock2008}, presented EyeWrite: a gaze typing system, where characters are entered by performing predefined gestures for each character.
EyeWrite achieves an average typing speed of 5 words per minute, and the participants felt it was easier to use EyeWrite than on-screen keyboard.
Bulling et al.~\cite{Bulling2009}, presented wearable EOG goggles using which gaze gestures can be performed as presented in~\cite{Drewes2007} to interact with computers.

Prior research that explores gaze- and foot-based multimodal interaction is limited for non-accurate secondary tasks and demonstrated for specific tasks like map navigation.
A generic gaze- and foot-based framework that can be used to interact with any application for point-and-click interactions on a computer is yet to be realized.
Similarly, the usability of gaze gestures are demonstrated for specific use cases like gaze typing or gaming, and also, the accuracy of gaze gestures is not much explored.
In this dissertation research, we propose solutions, frameworks, and results from evaluations that address majority of the concerns with gaze- and foot-based multimodal interaction, and gaze gesture-based interactions paradigm.

\section{Gaze and Foot Input Framework}
The basic principle behind gaze- and foot-based interaction framework is to achieve co-ordination between gaze and foot input to precisely point-and-click on the interface elements.
In this method, we use an eye tracker that tracks a user's gaze on the screen and a wearable device that is operated with the user's foot.
To click on a point of regard (button), the user first looks at the location and presses the pressure pad, with the foot, attached to the wearable device.
The foot input is translated into mouse clicks that are executed on the interface element that the user is currently looking at.
We have developed two systems that use the gaze- and foot-interaction framework.

\subsection{GAWSCHI}
\begin{figure}
\centering
\includegraphics[width=0.6\columnwidth]{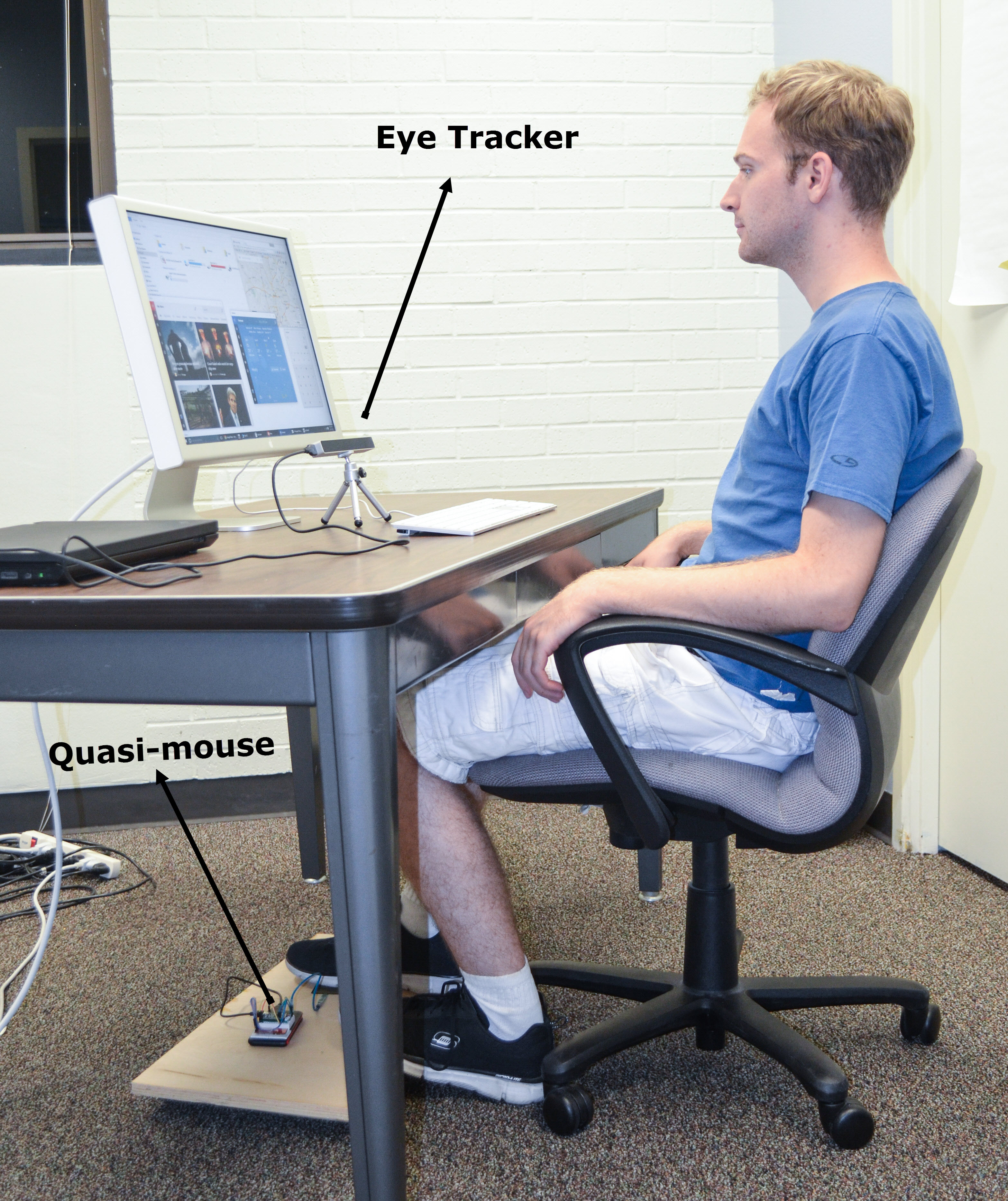}
\caption{A User Interacting with a Computer Using GAWSCHI}
\label{figure:gawschiuser}
\end{figure}

GAWSCHI, a Gaze-Augmented, Wearable-Supplemented Computer-Human Interaction framework that combines both gaze and foot input to effectively address the ``Midas Touch" problem (Figure \ref{figure:gawschiuser}).
GAWSCHI enables accurate and quick gaze-driven interactions in a desktop environment, while being completely immersive and hands free.
The system uses an eye tracker and a wearable device (quasi-mouse) that is operated with the user's foot, specifically the big toe.
The system evaluation showed that gaze-driven interaction using GAWSCHI is as good (time and precision) as mouse-based interaction in the majority of tasks.
Also, an analysis of NASA Task Load Index post-study survey showed that the participants experienced low mental, physical, and temporal demand; they also achieved a high performance.
Further details regarding GAWSCHI can be found at~\cite{Rajanna:2016}.

\textbf{Goals to be achieved:} GAWSCHI currently supports click, double-click, click-and-hold, and hold-and-release interactions.
However, interactions like pan, zoom, scroll, etc., are not yet directly supported.
To be coherent with user expectations, an interaction system should support mouse-like interactions.
Hence, to adopt GAWSCHI as a complete interaction system, it is required to implement all the interactions discussed above, and the interactions should feel natural.
Lastly, we need to understand the learning abilities of users for mouse- and foot-based interaction.

\subsection{Gaze Typing}
Gaze Typing is a gaze-assisted text entry method using an on-screen keyboard and gaze-input.
There are mainly two advantages of gaze typing: 1) it serves as a primary means of communication for those who encountered a serious injury resulting in the loss of speech and motor functions \cite{Gazetyping:twentyYears:Majaranta}, and 2) for users with no disability, gaze typing allows for hands-free text entry, hence supporting rich interactions.
Despite some crucial applications of gaze typing, this method is limited by its lower typing speed, higher error rate, and the resulting visual fatigue, since dwell-based key selection is used.

\begin{figure}[!ht]
\centering
\includegraphics[width=0.9\columnwidth]{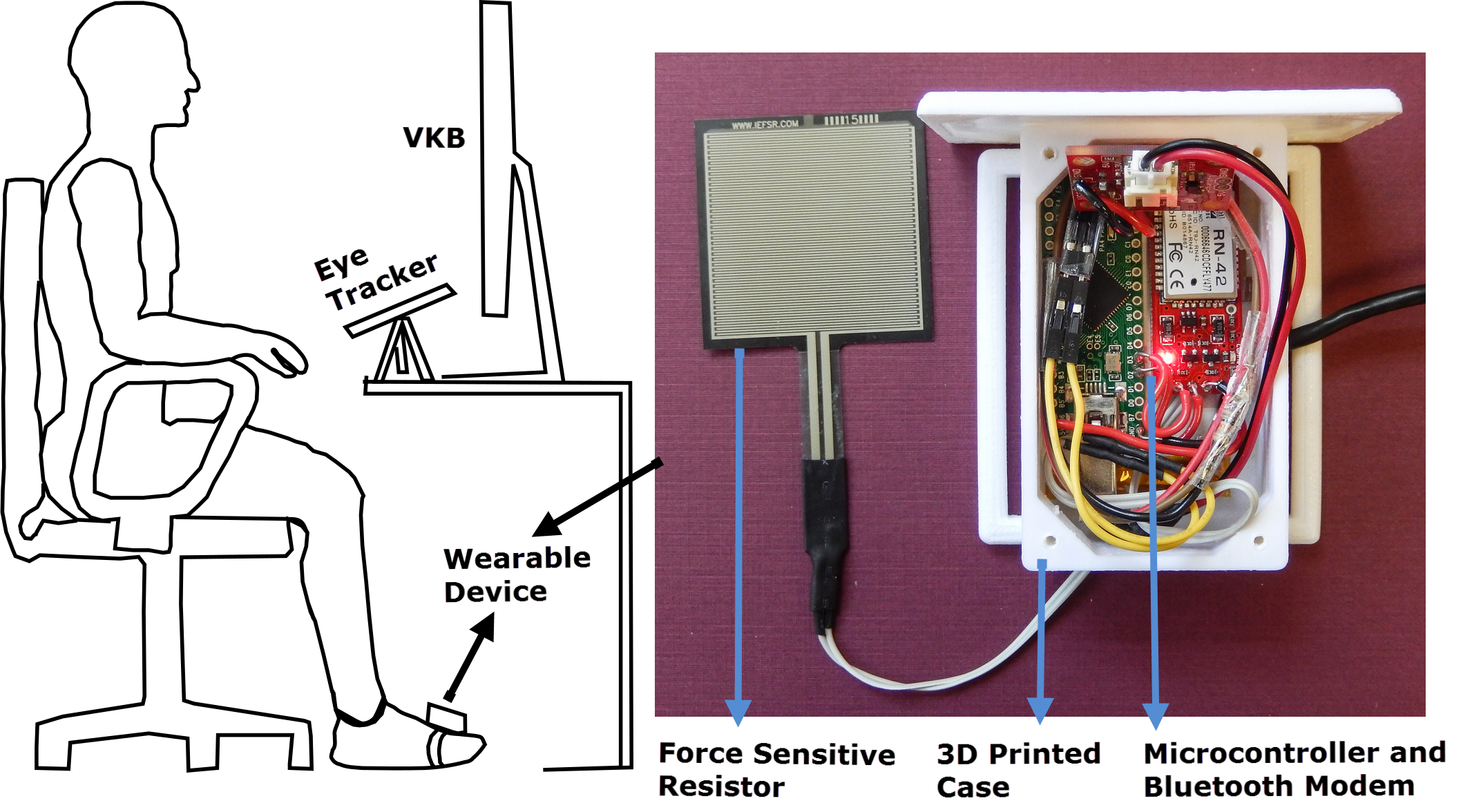}
\caption{Gaze typing system with an eye tracker and foot operated wearable device.}
\label{figure:gazetypinguser}
\end{figure}

\begin{figure}[!ht]
\centering
\includegraphics[width=0.8\columnwidth]{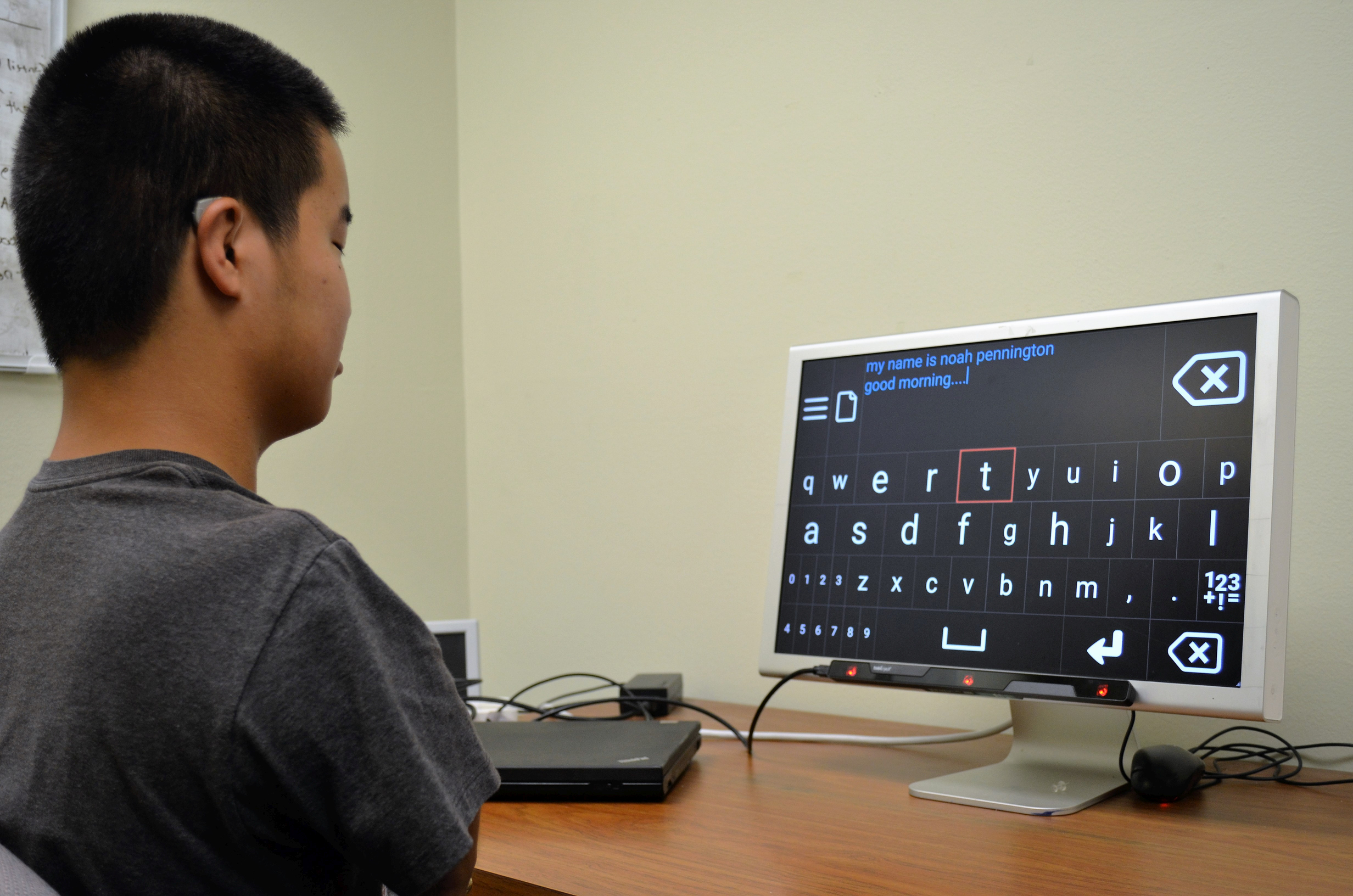}
\caption{A user with physical impairment is performing text entry using the foot-operated gaze typing system.}
\label{figure:gazetypinguserboy}
\end{figure}

We have developed a gaze-assisted, wearable-supplemented, foot interaction framework for dwell-free gaze typing (Figure~\ref{figure:gazetypinguser}). 
The framework consists of a custom-built virtual keyboard, an eye tracker, and a wearable device attached to the user's foot. 
To enter a character, the user looks at the character and selects it by pressing the pressure pad, attached to the wearable device, with the foot (Figure~\ref{figure:gazetypinguserboy}).
Results from a user study involving three participants with motor impairments showed that the participants achieved a mean gaze typing speed of 7.39 Words Per Minute (WPM); also, the mean value of Key Strokes Per Character (KPSC) was 1.06, and the mean value of Rate of Backspace Activation (RBA) was 6\%. 
Furthermore, from a comparative study involving three participants with no impairments, the participants achieved a mean typing speed of 10.48 WPM (typical 6.97), 1.09 KPSC (ideal 1.0), and 9\% RBA (ideal 0\%).
Further details regarding this work can be found at~\cite{Rajanna:gazetyping}.

\textbf{Goals to be achieved:} First, we would like to incorporate next word suggestion based on the last word entered and the current context.
Next, we need to understand how does varying the size of the on-screen keyboard and using different keyboard layouts impact speed and accuracy of gaze typing.
Lastly, we will further enhance the design and usability of foot-operated wearable device.

\section{Gaze Gesture Framework}
In this approach we re-contextualize gaze input as gestural input and each such gesture represents a piece of information based on the application context.
For example, a gaze gesture-based authentication system will use multiple gestures as a password, or in a gaze gesture-based interaction method, each gesture can represent commands like window minimize, maximize, play, pause, etc.
Generally, a gaze gesture consists of a series of fixation points connected with saccadic eye movements creating a scan-path (gaze-path).
Such a scan-path can be further processed using sketch recognition and pattern matching algorithms for specific use cases.
Based on this principle, we have developed two systems that leverage gaze gestures: 1) an authentication system, and 2) gaze gesture-based interaction paradigm.

\subsection{Gaze Gesture-Based Authentication}
\begin{figure}[!ht]
\centering
\includegraphics[width=0.9\columnwidth]{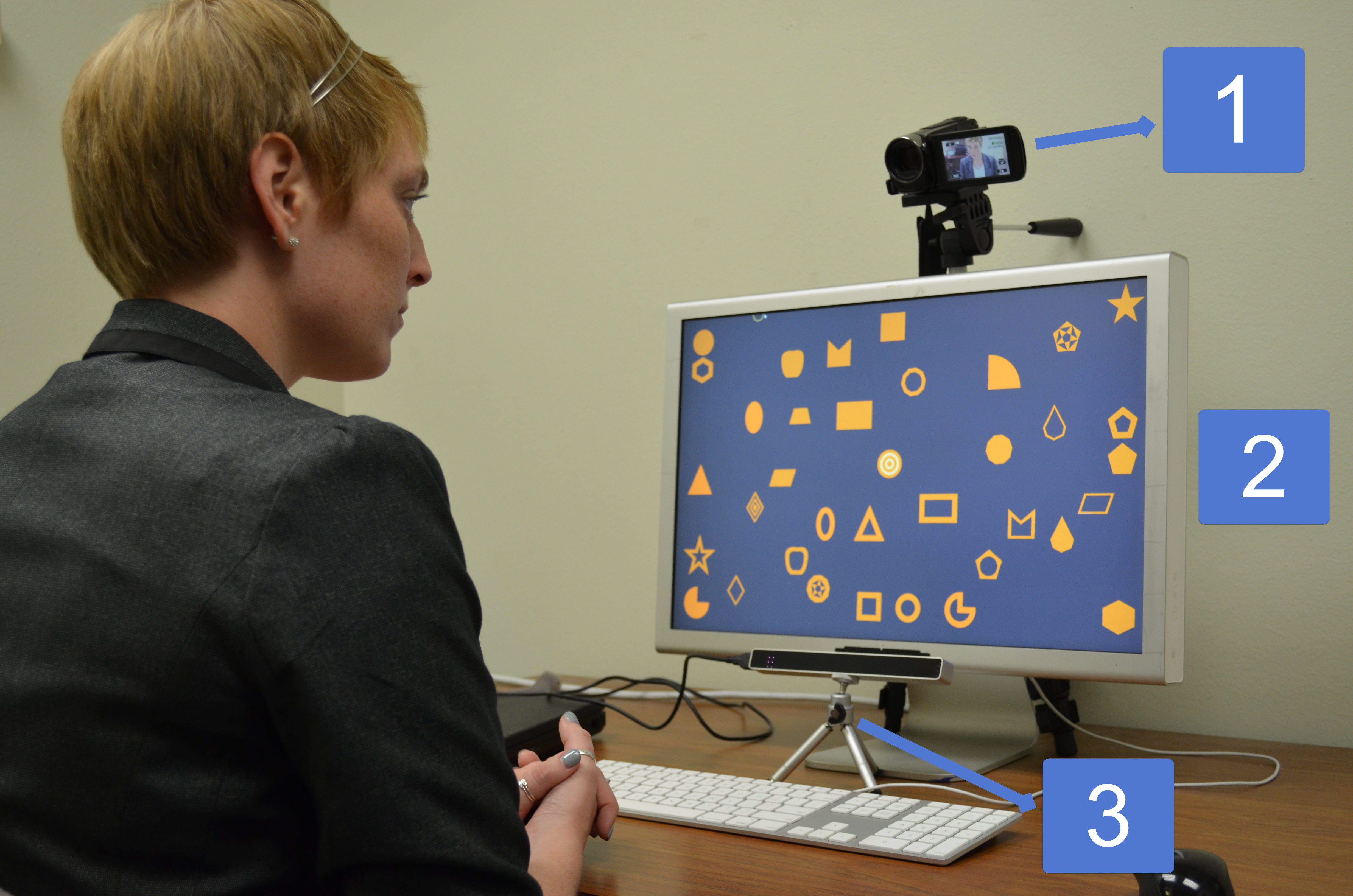}
\caption{Gaze gesture-based authentication [1-camera, 2-authentication interface, 3-eye tracker].}
\label{figure:gazeauthentication}
\end{figure}

Shoulder-surfing is the act of spying on an authorized user of a computer system with the malicious intent of gaining unauthorized access.
Current solutions to address shoulder-surfing such as graphical passwords, gaze input, tactile interfaces, and so on are limited by low accuracy, lack of precise gaze-input, and susceptibility to video analysis attack.
We developed an intelligent gaze gesture-based system that authenticates users from their unique gaze patterns onto moving geometric shapes~(Figure \ref{figure:gazeauthentication}).
The system authenticates the user by comparing their scan-path with each shapes' paths and recognizing the closest path.
In a study with 15 users, authentication accuracy was found to be~99\% with true calibration and~96\% with disturbed calibration.
Also, our system is~40\% less susceptible and nearly nine times more time-consuming to video analysis attacks compared to a gaze- and PIN-based authentication system.

\textbf{Goals to be achieved:} From our current work, we have identified several potential next steps for improvements and extensions.
Currently, our system takes 15 seconds to authenticate a user, however, we want to supplement our strong authentication accuracy with further reducing authentication time to less than 7 seconds.
Furthermore, we will be randomizing the movement of shapes to completely prevent video analysis attacks.
Lastly, we would like to further scale our system to accommodate smaller form-factor devices from our current desktop-setting assumptions.

\subsection{Gaze Gesture-Based Interaction Framework}
\begin{figure}[!ht]
\centering
\includegraphics[width=0.9\columnwidth]{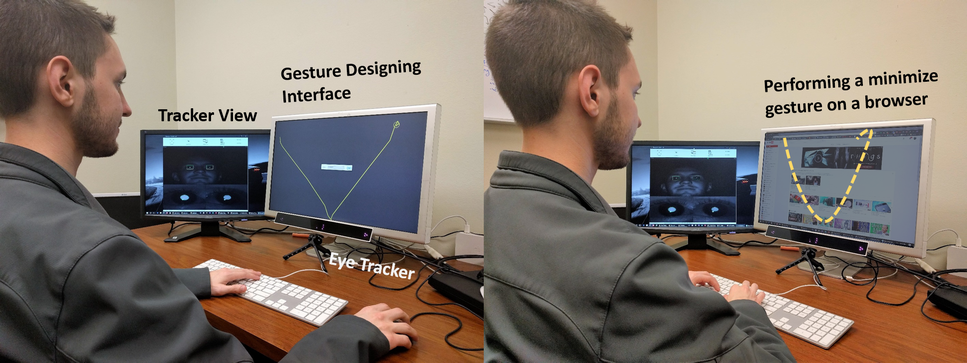}
\caption{A user creating a minimize gesture on the gesture training interface (left). The user executing the minimize gesture on a browser (right).}
\label{figure:gazegestureuser}
\end{figure}

We developed a gaze gesture-based interaction framework, where a user can interact with a wide range of applications using a common set of gestures (Figure~\ref{figure:gazegestureuser}).
In our framework, a gaze gesture can minimize, maximize, restore, or close an active window, or it can create a new tab, scroll, refresh, and so on a browser.
There are two significant advantages of our gaze gesture framework: 1) using gaze gestures, common interactions like minimize, maximize, scroll, and so on can be performed without switching the hand between keyboard and mouse, and 2) no need to remember a complex set of shortcuts, like shortcuts on a code editor, which vary across applications.
Additionally, we also preset a gesture training toolkit, where a user can create a new gesture and associate specific actions to it.
Results from a user study involving seven participants showed that the system accuracy was 93\% and the F-measure was 0.96.
We foresee this framework as an accessible solution to users with physical impairments and a rich interaction paradigm to non-disabled users.

\textbf{Goals to be achieved:} We will further focus on adding new gestures to the system, while keeping the accuracy high.
Also, we will explore various machine learning algorithms for gesture recognition to improve the recognition time over current template matching algorithm.


\section{Contributions}
In this research, we focused on integrating gaze modality for everyday human-computer interactions both as a rich and accessible interaction paradigm.
We approached this by developing a gaze and foot-operated multimodal framework to achieve point-and-click interactions on a computer, and a gaze gesture recognition framework to achieve gaze gesture-based interactions on a computer.
Following are some of the primary contributions of our research:
\begin{itemize}
    \item By developing GAWSCHI~\cite{Rajanna:2016} based on gaze- and foot-interaction, we demonstrated that a multimodal framework based on gaze modality can be used to achieve point-and-click interactions with an accuracy that is at least as good as a mouse in the majority of tasks.
    \item For point-and-click interactions, combing gaze with a supplemental input like foot input forms an effective solution to the Midas touch issue~\cite{Rajanna:2016}.
    \item When using gaze and foot-based multimodal framework for interacting with a computer, users experience low mental, physical, and temporal demand~\cite{Rajanna:2016}.
    \item By enabling gaze typing through foot-input modality, we have demonstrated a dwell-free gaze typing system~\cite{Rajanna:gazetyping}.
    \item We showed that the dwell-free, foot-operated gaze typing system, when evaluated by users with physical impairments, achieved a gaze typing speed of 7.39 words per minute (WPM), with significantly a low error rate of 6\%. Also, the gaze typing speed with able-bodied users was 10.48 WPM~\cite{Rajanna:gazetyping}.
    \item We demonstrated a novel approach to counter shoulder-surfing attacks by developing a gaze gesture-based authentication system~\cite{Rajanna:chi17}.
    \item The gaze gesture-based authentication system effectively addresses shoulder-surfing while achieving an authentication accuracy of 99\% with true calibration, and 96\% with disturbed calibration.
    \item Furthermore, our system is~40\% less susceptible and nearly nine times more time-consuming to video analysis attacks compared to a gaze- and PIN-based authentication system.
    \item We demonstrated the ability to interact with various applications on a computer through a common set of gestures by implementing a gaze-gesture interaction framework that recognizes gaze gestures with an accuracy of 93\% (F-measure 0.96).
    \item Using gaze-gesture training toolkit, a user can create new gestures and associate specific actions to each gesture. This enables a rich set of interactions.
\end{itemize}

\bibliographystyle{abbrv}
\bibliography{paper.bbl}  
\end{document}